\documentclass[review]{elsarticle}%

\journal{International Review of Economics and Finance}

\usepackage{graphicx}
\usepackage{amsmath}
\usepackage{amssymb}
\usepackage{epsfig}
\usepackage{arydshln}
\usepackage{mathtools}
\usepackage{xcolor}
\usepackage{moreverb}
\usepackage{caption}
\usepackage{subcaption}

\newtheorem{thm}{Theorem}

\newtheorem{rem}{Remark}

\newtheorem{ass}{Assumption}

\newcommand{\clm}{\color{black}}

\newcommand{\clgy}{\color{black}}

\begin{document}
\begin{frontmatter}

\title{Evaluation of catch-up paths by an uncertain dynamic game model}

\author[add1]{Ilona Cserh\'{a}ti}
 \ead{ilona.cserhati@uni-corvinus.hu}

\author[add2]{\'{E}va Gyurkovics}
 \ead{gye@math.bme.hu}

\author[add1]{Tibor Tak\'{a}cs}
 \ead{takacs.tibor@uni-corvinus.hu}

\address[add1]{Corvinus University of Budapest, 1093 Budapest, 8 F\H{o}v\'am t\'er}
\address[add2]{ Mathematical Institute, Budapest University of Technology and Economics, Budapest, Pf. 91,
1521, Hungary }

\begin{abstract}
To model the interaction of fiscal and monetary policy, a novel discrete-time, uncertain, infinite time horizon, dynamic game model is developed, where the uncertainties
of expectations are modeled by unknown nonlinear but quadratically constrained deterministic functions.
Cost-guaranteeing Nash strategies are defined for fiscal and monetary policy
as two players. The model is suitable for comparative analysis of the development paths of catching-up economies. Specifically, we evaluate nine possible development paths for the
Hungarian economy, where each path is characterised by a proxy for the debt-to-GDP ratio.
\end{abstract}


\end{frontmatter}

\section{Introduction}

\subsection{Independent vs. coordinated fiscal and monetary policy as a game}

The independence of fiscal and monetary policy refers to the ability of the central bank and the government to manage the economy through their autonomous economic management.
Alternatively, the government may have influence over interest rate policy, or they may carry out their tasks according to pre-agreed long term common objectives.
Whether independent or coordinated fiscal-monetary policy is more efficient has been widely debated in economic theory. In practice, both solutions can be observed. While some
countries are known for their high degree of central bank independence, where central banks can operate without direct political influence (Switzerland, Canada, Norway, Singapore),
other economies follow a different path where fiscal and monetary policy work closely together, especially in times of economic crisis (Argentina, Brazil, India, South Korea,
United States). Many economists in the economics literature argue that central banks should operate independently of political influence to maintain low inflation and promote
economic stability. This independence could also lead to more predictable monetary policy. Kydland and Prescott (1977), for example, indicate that, because of the problem of time
inconsistency, an independent monetary authority may be more helpful in controlling inflation than when monetary policy is subject to political pressure. Barro and Gordon (1983)
also put forward the idea that politicians may have incentives to create surprise inflation that serves short-term economic growth but undermines long-term economic stability.
Fischer (1996) analyses the rationale for central bank independence and discusses its benefits for the effectiveness of inflation control. On the other hand, the implementation
of coordinated fiscal and monetary policies also has advantages: coordination can contribute to the achievement of common economic objectives (inflation, growth, stability, etc.)
and provide a more comprehensive approach to economic challenges. Blinder (1997), for example, stresses the importance of central bank independence, but also the need for coordination
with fiscal policy, especially in times of crisis. Woodford (2003) provides a comprehensive theoretical framework for describing the interaction between monetary policy and fiscal policy,
emphasising the importance of a commitment to common long-term objectives. Masciandaro and Quintyn (2008) examine the implications of central bank independence and the necessary
interaction between monetary and fiscal policy, particularly in the context of global financial crises. Both independent and coordinated fiscal and monetary policy can be effective
depending on the economic, institutional and political environment of a country. Finding the right balance between independence and coordination is a key to optimising economic outcomes.

In practice, monetary policy has certain guarantees that it can make decisions independently of fiscal policy. This is particularly important in catching-up economies, where
governments often tend to prioritise growth considerations over equilibrium requirements. However, monetary policy must stick to its basic task of ensuring the stability of the currency.
Dynamic game theory is a useful tool to model the interaction between the two policies.

When in the 1970s the need for monetary policy to be at least partially independent of fiscal policy in order to fight inflation arose, a number of game-theoretic models were
developed to analyse the interactions between the two policies. The degree of independence varies across national economies and may change over time, see the historical overview of the
period 1972-2012 of Garriga (2016). Accordingly, game-theoretic analyses include both cooperative and non-cooperative games, often comparing the economic policy implications of the two basic models. The use of difference or differential games is an obvious way to represent the interactions and to monitor the macroeconomic effects of the two policies. Some of the models are formalised but
not computable, so they are essentially qualitative and can be used to draw theoretical conclusions. However, there are also computable game models that can be used to inform concrete
decisions. A relatively early paper is Tabellini (1986), which applied a continuous-time model to compare the cooperative and non-cooperative cases in terms of the evolution of public
debt and the speed of adjustment to a steady-state solution.  Petit (1989) used a quantified continuous-time linear-quadratic game model to make specific calculations for the Italian economy, comparing the evolution of the costs of two players in non-cooperative and cooperative games.  In addition to deterministic games, the application of stochastic game models also emerged. Leeper's (1991) model includes utility-maximising households alongside fiscal and monetary authorities.  The paper examines the stability properties of the dynamic model and the corresponding theoretical and policy
implications as a function of the control parameters of fiscal and monetary policy. The model is based on stochastic assumptions: the distribution of variables and their parameters are assumed to be known.   The dynamics of the system is formalized by a linearized model, and the equilibrium is interpreted in terms of deviations from the deterministic szeady state. Petit (1989) also uses sub-models for cooperative approaches in his model. The game model of Bassetto (2002) is based on a similar bargaining process. Here the two players are the household (private) sector and the
government (state); however, the control of fiscal and monetary policy is in one hand. The purpose of the model is to prove that under certain conditions the price level is
determined only by fiscal variables.
The analysis and modelling of the relationship between fiscal and monetary policy has been brought to the fore again, particularly in the context of the economic impact of the
two major recent crises, the global financial crisis that started in 2008 and the pandemic of 2020-2022. The models make use of new theoretical results and methods
developed in the field of dynamic systems and dynamic games.
Cui (2016), Stawska et al. (2019) used non-cooperative games to model the effects of the
interactions between the two policies. Feedback Nash strategies are determined e.g. in the paper Erickson (2011) for a continuous model without uncertainty.
J{\o}rgensen and Zaccour (2014) gave an overview of applied static and dynamic games on the area of cooperative advertising. Both cooperative and non-cooperative game models have been used by Saulo (2016), Bluescke et al. (2023), Saltari (2022), Nikooeinejad (2022), Engwerda et al. (2019).
Tabellini (1986), Saulo (2016) and Nikooeinejad (2022) gave Stackelberg solutions, the former with monetary policy and the latter with fiscal policy playing the leading role, thus
different models can be found in the literature.
This paper attempts to evaluate the catching-up paths of the Hungarian economy using a discrete-time dynamic game.
Taking into account the domestic experience, the authors opted for a non-cooperative model. Experience shows that the growth imperative characteristic of populist economic policies typically
leads to conflict between the two policies, with little coordination. Here Nash cost-guaranteeing equilibrium solutions for the two players are determined (see Gyurkovics Tak\'{a}cs (2024)),
and the development and catch-up paths are assessed and compared on the basis of the evolution of public debt. The reduction of public debt and keeping the debt-to-GDP ratio
below a certain percentage are both a criterion for joining the euro area and regulated by the Hungarian Constitution.

\subsection{Statement of the problem}

The objective of this paper is to develop a discrete-time, uncertain, dynamic, linear-quadratic game-theoretic model of the interaction between fiscal and monetary policy. Expectation variables 
are often
included in such models, notably expected inflation, which has a significant feedback effect on real variables. Since the variables of expectations in the dynamic equations are inherently 
uncertain, it is important to take uncertainties into account in the dynamic equations. To do so, one possible solution is to use stochastic game models, see e.g., Cui (2016), Saulo (2016) and
Nikooeinejad (2022). However, this approach requires such stochastic assumptions,
which are often not met in economic models.  Some papers model uncertainties by including external perturbations (Engwerda (2017), Kebriaei and Ianelli (2017), V\'{e}lez et al. (2021),
Gyurkovics and Tak\'{a}cs (2005)), eliminating the need for stochastic assumptions, and using different interpretations of the purpose of the game (min-max, hierarchical, etc). 
Parameter uncertainties were considered (Engwerda et al. (2019)), and the solution of a cooperative game was determined by a common
objective function.
In this paper, another robust approach is applied, which is also well suited for expectation modelling not requiring any  stochastic assumptions.
It differs from both the frequently applied adaptive and rational expectations models.

The uncertainties of the variables of expectations in the dynamic equations are modeled by unknown deterministic nonlinear 
functions, which satisfy certain constraints given in advance. In this way, the dynamic
equation of the game is not uniquely defined, since an infinite number of admissible functions describing the uncertainty can be realized. 
Therefore non-cooperating players cannot
optimize their objective functions, since there can be as many optima as there are possible realizations of the admissible uncertainty functions. However, under certain
conditions, it is possible for players to choose strategies that, though do not yield a minimum of their objective functions, will keep their objective functions
below a well-defined finite value for any possible model. These are called {\it cost guaranteeing strategies.}

In this robust approach, a relaxed notion of Nash
equilibrium can be interpreted, as well. In particular, a set of strategies will be called {\it  Nash cost-guaranteeing strategies} if, on the one hand, they represent a cost-guaranteeing
strategy in the sense described above and, on the other hand, there is an admissible uncertainty such that, under the corresponding model, these strategies provide a Nash
equilibrium solution in the traditional sense, i.e., there exist realizations of uncertainties that define this type of equilibrium. These are
strategies of players for which it is true that, if any player deviates from a given strategy while the other players continue to follow the Nash strategy, then the guaranteed
cost of the player choosing the deviating strategy will deteriorate. Here we content ourselves with this informal description of the notion of Nash guaranteeing strategies. For exact mathematical definition, we refer to the work of Gyurkovics and Tak\'{a}cs (2024).

In the authors' best knowledge, a similar approach has not yet been used in fiscal-monetary game models.
Nash cost-guaranteeing strategies are determined for fiscal and monetary policy as two players. In this fiscal-monetary game model, deviations from given reference trajectories
are penalized by quadratic functions over infinite time horizon. Due to uncertainties, it is not possible to determine the optima of quadratic functions, and instead
Nash cost-guaranteeing strategies are determined as equilibrium strategies for players.

The proposed model is suitable for evaluating and comparing the development paths of catching-up economies. Specifically, different catching-up paths are defined and evaluated for the
Hungarian economy. The resulting catching-up paths are characterised and compared by the evolution of a proxy for the debt-to-GDP ratio. The model is used to examine nine
catching-up scenarios for the Hungarian economy. Each catching-up scenario is constructed on the basis of the target catching-up path and the projected size of the central
budget deficit.
In the approach presented in this paper, expectations are always influenced
by the measured value in the previous period, but this can be modified by modelled uncertainties within certain limits. The contributions of the present paper are as follows.
\begin{itemize}
\item A theoretically new robust discrete-time non-cooperative linear quadratic dynamical game is proposed to model fiscal-monetary policy interactions over infinite time horizon.
\item In this model, new Nash cost guaranteeing strategies are applied introduced by the authors.
\item The uncertainties of the expectation variables are modelled non-stochastically with unknown quad\-ratically constrained nonlinear functions.
\item This approach provides a theoretically new way to model the expectation variables in dynamic economic models.
\item •	The proposed theoretically new, computable model can be used to analyze and compare catch-up paths of any emerging economy.
In particular, is it applied to evaluate alternative catch-up paths for the Hungarian economy.
\end{itemize}
The rest of the paper is organized as follows. The fiscal-monetary game model is set up in section 2. In section 3, the theoretical preliminaries are included.
Section 4 is a comparative analysis of the catching-up paths defined by the model, and Section 5 summarizes the conclusions.

\section{Setting up the uncertain game model}

For a game-theoretic model of the combined effects of fiscal and monetary policy, a model has to be defined that captures the dynamics of the main variables of the two policies.
Such a model is considered, where both policies are described by a single dynamic equation. Quadratic objective functions are defined for both players.
The constants of the dynamic equations were calculated on the basis of long time series using the ordinary least squares (OLS) method. Note that a simplified
version of the above model was used as a motivating example in Gyurkovics and Tak\'{a}cs (2024).

\subsection{The dynamic equations}

For the dynamic game model, the dynamic equations have to be defined that describe the interactions and temporal effects of fiscal
and monetary policy. One equation for each player is considered. These two dynamic equations can be seen as a modification of the model of Saulo et al. (2016).

Regarding the dynamics of the real sphere, it is supposed that for $t=1,2,...$, the planned
nominal path of the GDP $\xi_{t}^{*}$  and also the planned nominal values of the central budget balance $g_t^{*}$
are given.
Throughout the paper, the time series of variables indexed with an asterisk in each case represent a pre-specified desired reference path.
In this sense, reference, planned and target trajectories are synonymous.
Introduce the variable $z_t$ for the relative deviation from the reference path and let $g_t$ be the relative deviation
of the budget balance projected to the planned nominal GDP:
\begin{align*}
z_t = \frac{\xi_t -\xi_{t}^{*}}{\xi_{t}^{*}}, \hspace{1cm}  g_t=\frac{\overline{g}_t-g_t^{*}}{\xi_{t}^{*}},
\end{align*}
where $\xi_t$ is the actual nominal GDP, $\overline{g}_t$ is the actual nominal budget balance in period $t.$ The equation of the real sphere is
\begin{align}
z_{t+1}=-\alpha_1 \left(i_t - E\left[\pi_{t+1}\right]\right)-\alpha_2 g_t, \label{yuj1}
\end{align}
where $i_t$ is the nominal interest rate, $E[\pi (t+1)]$ is the expected inflation in period $t+1$, $\alpha_1$ and $\alpha_2$ are positive constants. According to (\ref{yuj1}), the
relative deviation from the reference path is negatively related to the real interest rate. Since the budget deficit increases GDP due to loose, demand-increasing fiscal policy,
the coefficient of the balance indicator $g_t$ is also negative (the surplus restrains the economic growth, while the deficit stimulates it).
The equation for the real sphere corrsponds to the IS curve of the model of Saulo (2016) for the deviation from the reference path. Already the variables in the dynamic
equation of Leeper's (1991) model were deviations from a reference trajectory, specifically from its steady state trajectory.  In that model, the author used a linear equation
approximating the dynamics. The method proposed in the present paper allows to keep the nonlinearities in the dynamic equations. An example is the expectation variable of
inflation in equation (3), the representation and interpretation of which is discussed in subsection 2.3.

Monetary policy sets an inflation target that it aims to achieve and maintain in order to preserve the value of the domestic currency. Denote $\pi^{*}$ this target level, which
is constant in time, and denote $i^{*}$ the corresponding interest rate level. The equation of the monetary policy is
\begin{align}
\pi_{t+1}&=\beta_ 1 E \left[z_{t+1}\right]+E\left[\pi_{t+1}\right]+\beta_2\left(i_t-i^{*}\right),   \label{yuj2}
\end{align}
where $E \left[z_{t+1}\right]$ is the uncertain expectation for the relative deviation from the planned GDP value.
The Phillips curve in Saulo (2016) has been modified by including the deviation of GDP from the reference path as an expectation variable in the present model, allowing
for non-linear effects (see subsection 2.3), and by including the effect of the deviation of interest rates from the reference path. In the model presented here, the
evolution of the debt ratio is not endogenous, but is considered as an observational variable with which different catch-up paths are evaluated.

\subsection{Objective functions of the players}

For each player, a quadratic objective function is defined over an infinite time horizon. The objective of fiscal policy is to keep growth and the evolution of the
budget balance close to the planning path:
\begin{align}
J_F=\sum_{t=0}^{\infty}\left( \gamma_1 z_t^2 + \gamma_2 g_t^2   \right) ,   \label{yuj3}
\end{align}
where $\gamma_1$ and $\gamma_2$ are positive constants. The fiscal policy intends to reach the lowest possible value of (\ref{yuj3}). At the same time, monetary policy aims to keep inflation and
interest rates close to the target:
\begin{align}
J_M=\sum_{t=0}^\infty \left( \varrho_1 \widetilde{\pi}_t^2 + \varrho_2  \widetilde{i}_t^2 \right) ,   \label{yuj4}
\end{align}
where $\widetilde{\pi}_t=\pi_t - \pi^* ,$ $\widetilde{i}_t=i_t - i^*,$ $\rho_1$ and $\rho_2$ are positive constants.
The monetary policy intends to reach the lowest possible value of (\ref{yuj4}).

\subsection{Uncertainty structure}

When modelling uncertain expectation $E[z_{t+1}]$ for the relative deviation of GDP from the target path, we assume uncertain but non-stochastic expectations instead of
the usual adaptive or rational expectations. This expectation is fundamentally influenced by, but not identical to, the actual deviation, but is assumed to be
\begin{align}
E\left[z_{t+1}\right]=z_t+p_1,   \label{yuj5}
\end{align}
where $p_1$ is an unknown function of time, of the relative deviation of GDP from the target path and of the inflation satisfying certain constraints with respect to the function values.  

When modelling inflation expectations, we also assume uncertain but non-stochastic expectations instead of the usual adaptive or rational expectation formulas. Expectation $E[\pi_{t+1}]$
of price increases are fundamentally influenced by the measured inflation in the previous year, but are not necessarily the same:
\begin{align}
E\left[\pi_{t+1}\right]=\pi_t+p_2,   \label{yuj6}
\end{align}
where $p_2$ is a nonlinear deterministic unknown function of time, of the relative deviation of GDP from the target path, and of the inflation, which is not assumed to have a specific shape,
but it satisfies certain constraints.

As far as the constraints are concerned, we choose them so that a wide range of uncertainties to be allowed: roughly speaking, we require that $p_1$ and $p_2$ take their values from cones similar to Figure 1 (a) and (b).
\begin{figure}[ht]
     \centering
             \includegraphics[width=9cm]{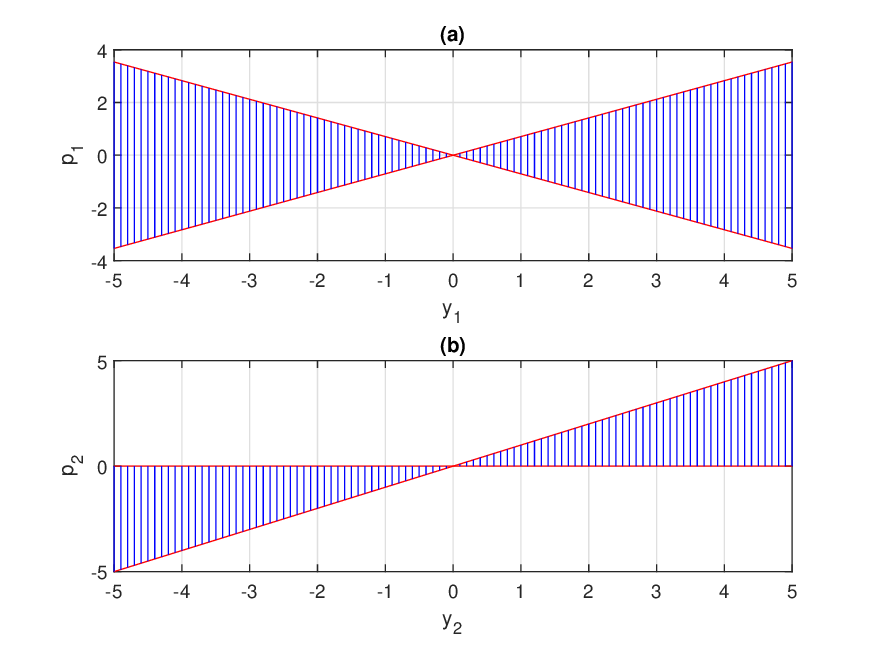} 
         \caption{{(a): the cone defined by (\ref{L1}),\hspace{0.5cm} (b): the cone defined by (\ref{L2}) }}
\end{figure}

The corresponding constraints can formally be given by equations
\begin{align}
& -\frac{1}{\sqrt{2}} \left|y_1\right|\leq p_1 \leq \frac{1}{\sqrt{2}}\left|y_1\right| , \label{L1} \\
& -\frac{1}{2}\left|y_2\right| +\frac{1}{2}y_2\leq p_2 \leq \frac{1}{2}\left|y_2\right|+\frac{1}{2}y_2 , \label{L2}
\end{align}
where we have taken for simplicity $y_1=\delta_{11}z+\delta_{22}\widetilde{\pi}, and $ and $y_2=\delta_{21}z+\delta_{22}\widetilde{\pi}.$ The symmetric cone (\ref{L1}) for $p_1$ seems to be suitable, but the statistical data of our specific application show that the asymmetric cone (\ref{L2}) for $p_2$ is more adequate. We note, however, that more general constraints are allowed by the theoretical results to be applied (see Subsection 2.4 and Section 3 below.)

\begin{rem} \label{rem:1000}
Several forms of uncertainties are admitted that satisfy the above relations. These inequalities are satisfied e.g. by the nonlinear functions
{\clgy{\
\begin{align*}
p_1=\left\{ \begin{array}{ccc}
               \frac{1}{\sqrt{2}}\left|y_1\right|\sin\left(\frac{1}{y_1}\right) & if & y_1 \neq 0 \\
              0 & if & y_1=0
            \end{array}
 \right.
\end{align*}
and
\begin{align*}
p_2=\left\{ \begin{array}{ccc}
             \frac{y_1}{2}+\frac{\left|y_2\right|}{2}\sin\left(\frac{1}{y_2}\right)
              & if & y_2 \neq 0 \\
              0 & if & y_2=0
            \end{array}
 \right.
\end{align*}
}}that have been applied in the simulations below. Moreover, the conditions allow a certain fluctuation of the coefficients as well.
If, e.g. $p_2=c_{2t}\widetilde{\pi}_t,$ with $0\leq c_{2t}\leq \overline{c}_2,$ then in case of $y_2=\overline{c}_2 \widetilde{\pi} $ inequality $0\leq p_2\leq\overline{c}_2\widetilde{\pi}$ holds true
if $0\leq \widetilde{\pi}$ and $\overline{c}_2\widetilde{\pi}\leq p_2 \leq 0,$ is satisfied if $\widetilde{\pi}\leq 0,$ i.e. inequality (\ref{L2}) holds true.
This means that on the right hand side of (\ref{yuj1}) the coefficient of
$\widetilde{\pi}$ can be $\alpha_1 \left(1+c_{2t}\right),$ while it can be $1+\beta_1 c_{2t}$ in (\ref{yuj2}). If $p_1=c_{1t}z_t,$ where $0\leq \left|c_{1t}\right|\leq \frac{\overline{c}_2}{\sqrt{2}},$
then in case of $y_1=\overline{c}_1 z,$ inequality $\left|p_1\right|\leq \frac{\left|y_1\right|}{\sqrt{2}}$ i.e. inequality (\ref{L1}) holds true. In this case the coefficient of $z_t$ in (\ref{yuj2}) can be
$\beta_1+c_1 sgn(z_t).$
\end{rem}

In summary, plugging (\ref{yuj5}) and (\ref{yuj6}) into (\ref{yuj1}) and (\ref{yuj2}), and using relation $\pi^* = i^*,$ one obtains the state-space representation
\begin{align}
z_{t+1} &= \alpha_1 \widetilde{\pi}_t -\alpha_1 i_t -\alpha_2 g_t +\alpha_1 p_2 \left(t,z_t,\widetilde{\pi}_t\right) \label{yuj7}  \\
\widetilde{\pi}_{t+1} &= \beta_1 z_t +  \widetilde{\pi}_t + \beta_2  \widetilde{i }_t + p_1 \left(t,z_t,\widetilde{\pi}_t\right)+p_2 \left(t,z_t,\widetilde{\pi}_t\right) . \label{yuj8}
\end{align}

\subsection{Canonical form of the model}
One can easily see that, with a suitable choice of variables, the above described model belongs to the class of two-players uncertain dynamic games, the canonical form of which can be given as follows: 

\begin{align}
 x_{t+1} &=Ax_{t}+B_1 u^1_{t}+ B_2 u^2_{t}+H p_{t},  \label{ff1} \\
q_{t} &= A_{q}x_{t}+G p_{t}, \label{ff3}  
\end{align}
where $x\in \mathbf{R}^{n_{x}}$ is the state, $u^1\in \mathbf{R}^{n_{u^{1}}}$ and
$u^2\in \mathbf{R}^{n_{u^{2}}}$
are
the inputs of Player 1 and  Player 2, $A, B_1, B_2, H, A_q$ and $G$ are given matrices of appropriate dimension.

The objective functionals are given as
\begin{align}\label{ff4}
  J_i(x_0,\bf{u}^1,\bf{u}^2) & =
 \sum_{t=0}^\infty \left(
x_t^TQ_i x_t+{{\clm u^1_t}}^{T} R_{i1}u^1_t+{u^2_t}^{T} R_{i2}u^2_t\right), \hspace{1cm} i=1,2,
\end{align}
with  matrices $Q_i=Q_i^T\geq 0$ and $R_{ii}=R_{ii}^T > 0,  \, R_{ij}=R_{ij}^T \geq 0, \,  i,j=1,2, \, i\neq j.$

All system nonlinearities/un\-certainties are represented by
function $p$
possibly depending on $t$ and $x$. Function $q$ is the
uncertain output. The only available information about $
p \in \mathbf{R}^{n_{p}}$ and
$ q \in \mathbf{R}^{n_{q}}$ is that
their values are constrained by the set
$\Omega = \Omega _{1} \times ... \times \Omega _{s}$,
\begin{align}
\Omega _{i} &=\left\{ \begin{bmatrix}
p_i \\
q_i
\end{bmatrix}
\in \mathbf{R}^{{\clm n_{p_i}+n_{q_i}}}:
\begin{bmatrix}
p_i \\
q_i
\end{bmatrix} ^{T}
\begin{bmatrix}
Q_{0i} & S_{0i} \\
S_{0i}^{T} & R_{0i}
\end{bmatrix}
 \begin{bmatrix}
p_i \\
q_i
\end{bmatrix} \geq 0
\right\} ,\hspace{1cm} i=1,...,s, \label{ff5}
\end{align}
where $Q_{0i}=Q_{0i}^{T}$, $R_{0i}=R_{0i}^{T}$ and $S_{0i}$ are constant
matrices, $p$, and
$q$ are partitioned appropriately.

The sets (\ref{ff5}) quadratically constrain the uncertainties/nonlinearities.
Specially, one obtains the model described in subsections 2.1-2.3 under the following choice of variables: $n_x=2,$ $n_{u^{1}}=1,$ $n_{u^{2}}=1,$ $s=2,$
\begin{align*}
  x_t & = \begin{bmatrix} z_t & \widetilde{\pi }_t \end{bmatrix}^T, \hspace{01.2cm} u^1_t= g_t, \hspace{1.6cm} u^2_t=\widetilde{i}_t, \\
  A & = \begin{bmatrix}0 & \alpha_1\\ \beta_1 & 1    \end{bmatrix}, \hspace{0.5cm}
   B_1 = \begin{bmatrix} -\alpha_2 \\ 0  \end{bmatrix},   \hspace{0.9cm}
   B_2 = \begin{bmatrix} -\alpha_1 \\ \beta_2  \end{bmatrix}, \hspace{0.5cm}
     H = \begin{bmatrix}0 & \alpha_1\\ \beta_1 & 1    \end{bmatrix},
\end{align*}
\begin{align*}       
    Q_1 & = \begin{bmatrix} \gamma_1 & 0 \\ 0& 0  \end{bmatrix}, \hspace{0.8cm}
   Q_2 = \begin{bmatrix}0 & 0 \\ 0 & \varrho _1 \end{bmatrix}, \hspace{0.5cm}
   R_1 = \gamma_2,  \hspace{1.0cm}
   R_2 = \varrho_2,\\
    A_{q_1} & = \begin{bmatrix} \delta_1 & \delta_2  \end{bmatrix}, \hspace{0.65cm}
   A_{q_2} = \begin{bmatrix} \delta_3 & \delta_4  \end{bmatrix}, \hspace{0.65cm} G_1=G_2=-1\\
Q_{01} & =-1,\hspace{0.2cm}  S_{01}=1, \hspace{0.2cm} R_{01}=1, \hspace{0.5cm} Q_{02}=0, \hspace{0.2cm} S_{02}=1, \hspace{0.2cm} R_{01}=0.
\end{align*}

\subsection{Estimation of the coefficients}
The constants in (\ref{yuj1}) and (\ref{yuj2}), which are specific to the Hungarian economy, were determined using the ordinary least squares (OLS) method based on time series data.
In (\ref{yuj1}), the constants were estimated on the basis of time series of the output gap (instead of $z_t$) calculated for the economy, and of corporate lending rates in the case of
interest rates ($i_t$). We used 20-year time series for the calculations, but excluded the two crisis periods 2008-2009 and 2020-2021. This also means that, for the future trajectories
simulated by the model, it is assumed that crises such as the financial or pandemic crises do not occur. The estimated coefficients of (\ref{yuj1}) and (\ref{yuj2}) are
\begin{align*}
\alpha_1=0.16, \; \alpha_2=0.19, \; \beta_1=0.699, \; \beta_2=0.433,
\end{align*}
while the coefficient of the objective functions (\ref{yuj3}), (\ref{yuj4}) and that of the uncertain outputs $y_1,$ $y_2$ are given as follows:
\begin{align*}
&\gamma_1=\rho_1= 0.2, \; \gamma_2=0.075, \; \rho_2=0.01, \\
&\delta_1=0, \; \;  \delta_2= 0.1, \; \; \delta_3= \delta_4= 0.15.
\end{align*}

\section{Preliminaries}
The theoretical background to the analysis is summarised below (see details in Gyurkovics and Tak\'{a}cs (2024)).

Introduce the following notations:
\begin{align*}
Q_0=\mbox{diag}\{ Q_{01},...,Q_{0s} \} , R_0=\mbox{diag}\{ R_{01},...,R_{0s} \} ,
S_0=\mbox{diag}\{ _{01},...,S_{0s} \} .
\end{align*}
For any positive constants $\omega _{j}$, $j=1,...,s$
 set
\begin{align*}
\underline{\omega} =\mbox{diag}\left\{ \omega _{1}I_{{\clm{n}}_{p_{1}}},...,\omega
_{s}I_{{\clm{n}}_{p_{s}}}\right\} , \hspace{1cm}
\underline{\underline{\omega}}=\mbox{diag}
\left\{ \omega _{1}I_{{\clm{n}}_{q_{1}}},...,\omega _{s}I_{{\clm{n}}_{q_{s}}}\right\}.  
\end{align*}
Furthermore, for any positive numbers $\tau_{j}^i$ and $\mu_{j}^i$ ($i=1,2,$ $j=1,...s$), set $\mathcal{S}_0 = S_0^T+R_0G,$ and
\begin{align}
 \Xi(\underline{\tau}^i,\underline{\underline{\mu}}^i) & = \underline{\tau}^i\left(\Xi_0 +
        \mathcal{S}_0 ^T \underline{\underline{\mu}}^i \mathcal{S}_0  \right)  \label{ff12}
\end{align}
\begin{ass} \label{ass:1}
Inequalities $R_0  \geq 0,$ and $\Xi_0 \doteq Q_{0}+{G}^{T}S_{0}^{T} +S_{0}  {G} + {G}^{T}R_{0}{G}<0 $ hold true.
\end{ass}

\begin{ass} \label{ass:2}
The matrix pair $\left(A, \left[B_1,B_2 \right]  \right)$ is stabilizable, and the matrix pair $\left(A, Q_1+Q_2  \right)$ is detectable.
\end{ass}

\begin{thm} \label{cor:3}
Suppose that the Assumptions 1,2 hold true.
If $ \widetilde{P}_i=\widetilde{P}_i^T>0,$ $\tau_j^i>0,$ $\nu_j^i>0,$ ($i=1,2, \; j=1,\ldots,s$), $K_1$ and $K_2$  satisfy for $i=1,2$ matrix
equation
\begin{align}                         \\
& \hspace{0.3cm}  \begin{bmatrix}R_{11}+ B_1^T \widetilde{P}_1 B_1 & B_1^T \widetilde{P}_1 B_2  \\
    B_2^T \widetilde{P}_2 B_1 & R_{22}+B_2^T \widetilde{P}_2 B_2\end{bmatrix}  \begin{bmatrix}K_1 \\K_2 \end{bmatrix}  = -  \begin{bmatrix}B_1^T \widetilde{P}_1 \\ B_2^T \widetilde{P}_2 \end{bmatrix}A,
    \hspace{1cm}
   \label{ff92}
\end{align}
and matrix inequalities
\begin{align}
& \hspace{3cm} 0  > \underline{\nu}^i  \Xi_0 +  {\underline{\tau}}^i \mathcal{S}_0 ^T \mathcal{S}_0   \label{ff103}\\
   0 & \geq
  \begin{bmatrix}
 \widetilde{\Phi}_i-\widetilde{P}_i &  \mathcal{A}^T \widetilde{P}_i & \widetilde{P}_i H & K_i^T & 0\\
 \widetilde{P}_i \mathcal{A} &   -\widetilde{P}_i   &  0 & 0 & 0\\
 H^T \widetilde{P}_i  &  0  & - H^T \widetilde{P}_i H + \underline{\tau}^i \Xi_0 & 0 & \mathcal{S}_0^T \underline{\underline{\tau}}^i\\
 K_i & 0 & 0 & -R_{ii}^{-1} & 0 \\
 0  & 0 & \underline{\underline{\tau}}^i \mathcal{S}_0 & 0 & - \underline{\underline{\nu}}^i
 \end{bmatrix}
 \label{ff102}
 \end{align}
where
$
\widetilde{\Phi}_i=Q_i + K_{\hat{i}}^T R_{i{\hat{i}}}K_{\hat{i}} +
A_q^T \left(\underline{\underline{\tau}}^i R_0+ \underline{\underline{\nu}}^i \right)A_q ,
$,
then
$\alpha_1(x)=K_1x$ and $\alpha_2(x)=K_2x$  are
Nash guaranteeing feedback strategies with approximate Nash guaranteed costs $V^i(x_0)=x_0^T P_i x_0, \; i=1,2$ where
$P_i$ given by $\widetilde{P}_i  =\left(P_i^{-1}+  H\Xi(\underline{\tau}^i,\underline{\underline{\mu}}^i)^{-1} H^T\right)^{-1}$ .
\end{thm}

The statement of the theorem follows from Corollary 4 of Gyurkovics and Tak\'{a}cs (2024).

\section{Analysis of the scenarios}

For the analysis of different scenarios, the results given in section 3 were applied to the model described in section 2. The initial values have been chosen 
as $x_0=[-0.04 \; 0.175]^T$. For these initial values, one obtains the approximate Nash guaranteed costs $V^1(x_0)=0.0117$ and $V^2(x_0)=0.0130$.
\begin{figure}[ht]
     \centering
             \includegraphics[width=9cm]{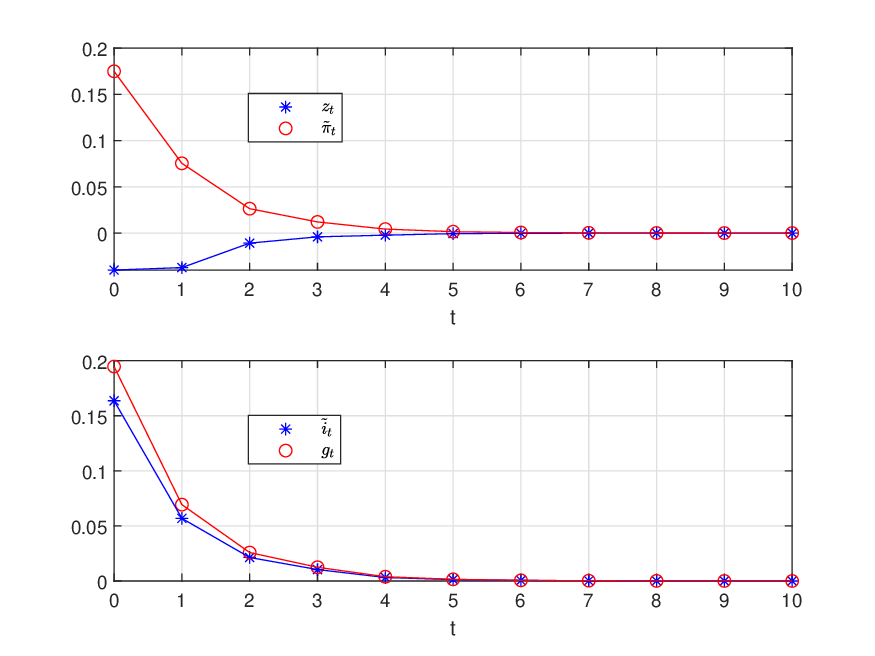} 
         \caption{{State and control trajectories of the game.}}
\end{figure}
Figure 2 shows that the deviations from the nominal values disappear rapidly.

Each scenario was evaluated by the evolution of the following indicator:
\begin{align}
d_t = \frac{D_t}{\xi_t},   \label{yuj77}
\end{align}
where
\begin{align}
D_{t+1} = D_t-\overline{g}_t,  \label{yuj88}
\end{align}
and $D_0$ is the state debt in the starting year of the simulation, i.e. in 2023. According to (\ref{yuj88}), a positive budget balance reduces the debt stock, a negative balance increases it.
Note that the interest paid on the previous debt stock is already included in $\overline{g}_t.$  Rate (\ref{yuj77}) is considered as a proxy for the debt-to-GDP ratio. 
We note that $d_t$ is not the actual
government debt-to-GDP ratio, as it does not model the effects of the maturity of past debt and possible debt swaps, which are also dependent on changes in external and internal interest
rates and exchange rates. Three target paths were defined for both growth and deficit, and all possible combinations of these paths gave the nine scenarios considered.
The following variants of nominal GDP growth $\xi_t^*$ were considered:
\begin{itemize}
  \item [] (1) Moderate growth, annually 2.575\%.
  \item[] (2) Average growth, annually 3.605\%.
  \item[] (3) Strong growth, annually 5.15\%.
\end{itemize}
Note that with a 3\% GDP deflator, the above scenarios correspond to annual volume growth of 2.5\%, 3.5\% and 5\% respectively. For $g_t^{\ast},$ i.e. for
the target budget deficit per nominal
target GDP, the following three variants were considered:
\begin{itemize}
  \item [] (A) A tight fiscal policy: 3\% in the starting year, then half a percentage point per year, and 0\% throughout once a balanced budget is achieved.
  \item[] (B) A loose fiscal policy: 6.7\%, 4.8\%, 3.5\% from the starting year, and 3\% for the rest of the years.
  \item[] (C) Populist fiscal policy: the same as (B), but for each election year (every fourth year from 2026) we have assumed 4.5\% instead of 3\%.
\end{itemize}
The scenarios (1A, 1B, 1C, see Figure 3) corresponding to the three growth dynamics under tight fiscal policy are similar in that the central budget balance turns zero in the fifth year and the
current inflation target of 3$\pm$1\% set by the Hungarian central bank is already achieved in the third year.
\begin{figure}[ht]
     \centering
             \includegraphics[width=9cm]{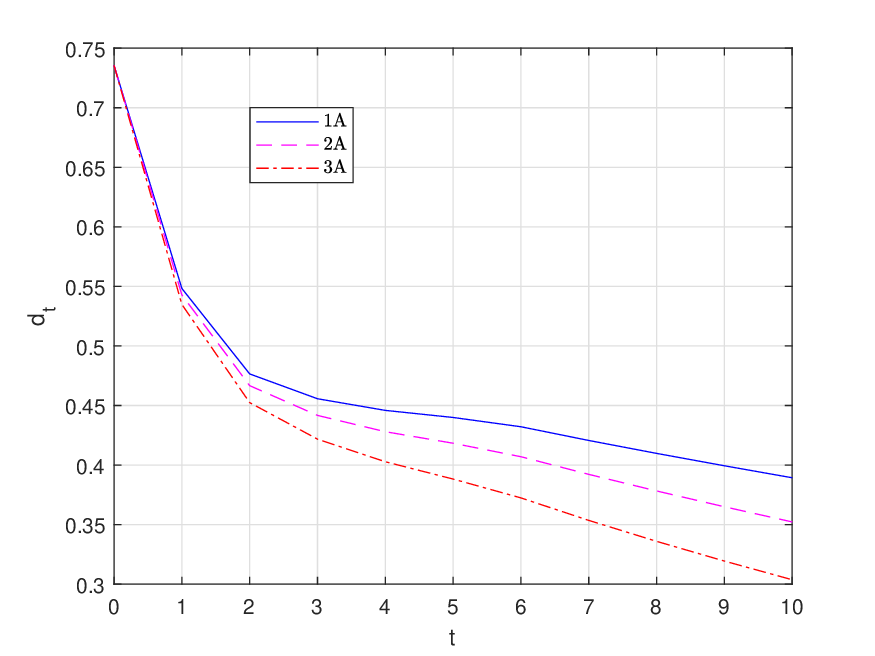} 
         \caption{{\clgy{Trajectories of $d_t$ with a tight fiscal policy }}}
\end{figure}
The debt ratio $d_t$ falls below 50\% in two years and remains on a downward trend
throughout. It makes sense that the more dynamic the growth, the more favourable the debt ratio run-off. These trajectories are sustainable in terms of fiscal debt and debt-to-GDP ratios.
Under the assumed tight fiscal path, of course, only moderate growth in the first place is realistic, assuming that a relatively rapid rebalancing of the central budget is feasible, growth
could be accelerated mainly by a noticeable increase in productivity in the short term. The mobilisation and proper use of external resources (capital transfers from the budget of the
European Union, foreign direct investment in high value-added activities) is essential for catching up.
The Hungarian experience shows that, due to mostly populist policies and economic governance, a fully balanced central budget is not a realistic scenario. A loose fiscal path assumes
that the central budget deficit as a share of GDP does not exceed 3\%, in line with the corresponding Maastricht criterion (the Maastricht criteria must be fulfilled by a country that wishes
to join the eurozone).
The criterion on the budget is in fact a balancing one, allowing a maximum of $\pm$3\% deviation in order to take account of the state of the economic cycle and the need for counter-cyclical
economic policies. However, populist policies, which in any case prioritise growth, tend to interpret this as keeping the deficit at or near the upper limit.
The paths for loose fiscal policy (1B, 2B, 3B, see Figure 4) have in common only in that they reach the inflation target in the third year.
\begin{figure}[ht]
     \centering
             \includegraphics[width=9cm]{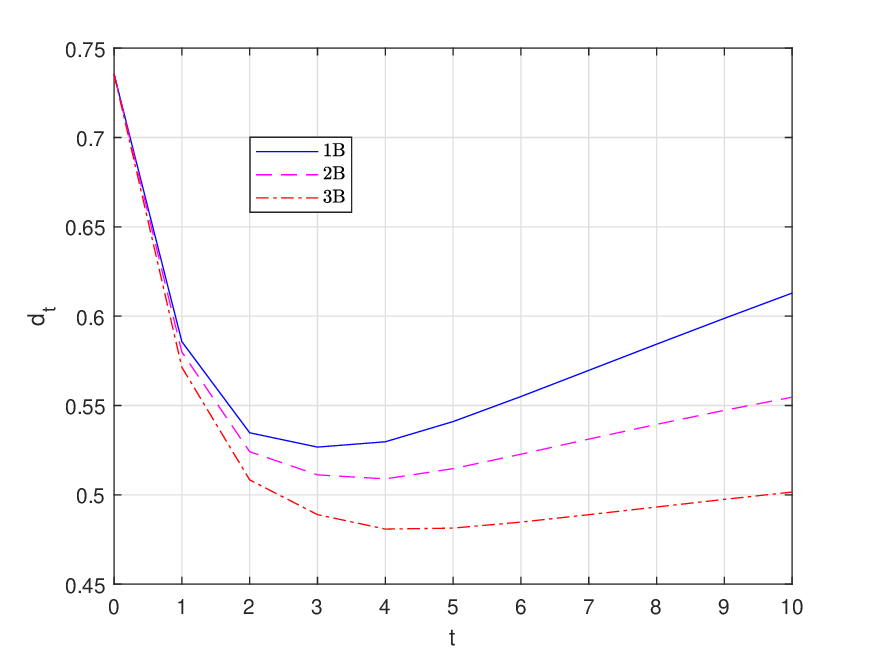} 
         \caption{{\clgy{Trajectories of $d_t$ with a loose fiscal policy }}}
\end{figure}
The paths 1B and 2B reach the 3\% of GDP
deficit in three years, the path 3B only in four years. In terms of debt ratios, paths 1B and 2B are not sustainable because the ratio is clearly on an upward path in the short term.
For path 3B, which corresponds to a high growth rate, the debt ratio stabilises around 50\%. In principle, this meets the Maastricht and Hungarian Constitutional criteria, but from an economic point
of view it is desirable to have a much lower rate. Note that three of the EU's catching-up economies (Lithuania, Estonia, Bulgaria) have managed to bring their debt ratios down to 40\% or below
after recovering from the pandemic crisis. The scenarios for the populist fiscal paths (1C, 2C, 3C, see Figure 5) are based on a budget deficit of 3\% of GDP in four years, with the inflation
target being met in the third year. Scenarios 1C and 2C are not sustainable because debt is on the rise. In the case of 2C, it takes slightly longer to return to the original value. For 3C, there
is no return to the starting point, but the rate stabilises slightly above 50\%, which is also not sustainable.
\begin{figure}[ht]
     \centering
             \includegraphics[width=9cm]{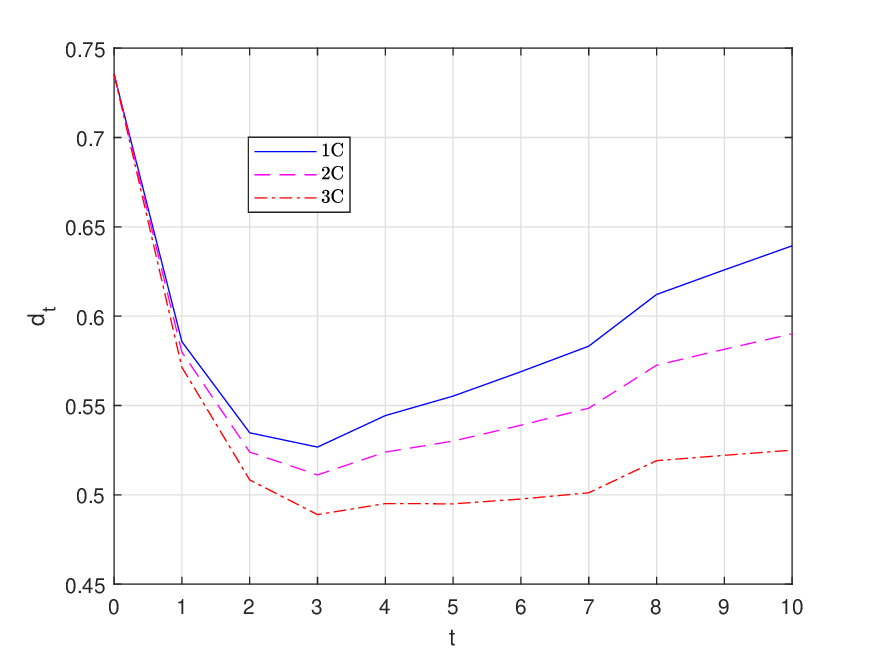} 
         \caption{{\clgy{Trajectories of $d_t$ with populist fiscal policy }}}
\end{figure}
The results of the model calculations therefore suggest that, in principle, fiscal austerity would be necessary for sustainability, which would require a significant change in the
Hungarian economic policy stance (paths 1A, 2A, 3A). Experience shows that fiscal austerity can only be enforced primarily by the EU rules and the threat of excessive deficit procedures. It is
unlikely that the central budget will be balanced even in the medium term. Even stabilising at a deficit of 3\% will only lead to a sustainable path if it is supported by productivity growth (path 3B).

\section{Conclusions}

We provided a new, previously unused method for modelling the interactions between fiscal and monetary policy: the method was applied to compare and analyze possible catching-up paths in Hungary.
The two policies were modeled by a discrete-time dynamic game, where the uncertainties of the expectation variables were represented by unknown, deterministic, quadratically constrained functions.
We defined Nash cost-guaranteed state feedback strategies for the two players, namely fiscal and monetary policy.
The catching-up trajectories under the given strategies are characterised by a proxy indicator for debt-to-GDP. The results show that one can only find sustainable catching-up scenarios only
under strict fiscal paths, i.e. under short-term fiscal balancing.
This would require a major change of mindset from the demand-expanding, growth-pushing populist economic policies. The Maastricht criterion was still just met even by a loose fiscal stance, with the
deficit remaining at 3\% in the longer term. This is still broadly acceptable, but the results suggest that the debt-to-GDP ratio is stabilising around the 50\% maximum allowed by the Hungarian Constitution. The debt rate, however, should be further diminished.

A further development of the model is the inclusion of public debt in the dynamic equations. The proposed method could also model additional uncertainties, such as the uncertainty in the
coefficients of the equations, but this would require the solution of some of additional technical problems.

\bibliographystyle{agsm}

\end{document}